\documentclass[11pt]{article}
\usepackage[top=2.0cm,bottom=2.0cm,left=2.54cm,right=2.54cm]{geometry}
\usepackage{mathrsfs,amsmath,amsfonts,mathtools,amssymb,wasysym,graphicx,subfigure,caption,rotating,comment,booktabs,cite}
\usepackage{bm}
\usepackage{color}  
\usepackage{extarrows}

\usepackage[hyperindex=true,
          pdfstartview=FitH,
          bookmarksnumbered=true,
          bookmarksopen=true,
          citecolor=blue,
          linkcolor=blue,
          colorlinks=true,
          pdfborder=001,
          unicode]{hyperref}

\allowdisplaybreaks
\DeclareMathOperator{\e}{e}
\newcommand{\D}{{\rm d}}

\renewcommand\({\left(}
\renewcommand\){\right)}

\newcommand{\blue}[1]{\textcolor{blue}{#1}}
\renewcommand{\blue}[1]{\textcolor{black}{#1}}

\newcommand{\fa}{\mathfrak{a}}
\newcommand{\fb}{\mathfrak{b}}

\newcommand{\re}{\mathrm e}
\newcommand{\pfrac}[2]{\frac{\partial #1}{\partial #2}}

\begin{document}
\title{General relativistic stochastic thermodynamics}

\author{
Tao Wang\thanks{
{\em email}: \href{mailto:taowang@mail.nankai.edu.cn}{taowang@mail.nankai.edu.cn}},~
Yifan Cai\thanks{{\em email}: \href{mailto:caiyifan@mail.nankai.edu.cn}
{caiyifan@mail.nankai.edu.cn}},~
Long Cui\thanks{{\em email}: \href{mailto:cuilong@mail.nankai.edu.cn}
{cuilong@mail.nankai.edu.cn}},~
and Liu Zhao\thanks{Corresponding author, {\em email}: 
\href{mailto:lzhao@nankai.edu.cn}{lzhao@nankai.edu.cn}}\\
School of Physics, Nankai University, Tianjin 300071, China}

\date{}
\maketitle
\begin{abstract}
Based on the recent work \cite{cai2023relativisticI,cai2023relativisticII}, 
we formulate the first law and the second law of stochastic thermodynamics in the 
framework of general relativity. These laws are established for
a charged Brownian particle moving in a heat reservoir and subjecting to an 
external electromagnetic field in generic stationary spacetime background, 
and in order to maintain general covariance, they are presented respectively 
in terms of the divergences of the energy current and the entropy density current.
The stability of the equilibrium state is also analyzed. 
\vspace{1em}

\end{abstract}

\section{Introduction}
\label{Introduction}

\blue{
With the advancement of nanotechnology, there is a growing need for a new 
thermodynamics framework which is applicable to mesoscopic systems, where 
heat exchange and energy fluctuations are nearly at the same order of magnitude,  
rendering traditional macroscopic thermodynamics inapplicable. By analyzing 
Langevin equation, Sekimoto \cite{sekimoto1998langevin} separated the energy changes 
into work and heat at the single-particle level, making the development of 
mesoscopic thermodynamics feasible. Sekimoto later named his theoretical framework 
\emph{stochastic energetics} \cite{sekimoto2010stochastic}. Stochastic energetics 
is based on Langevin equation and the corresponding Fokker-Planck equation (FPE), 
and the elegance of this theory lies in its seamless establishment of the relationship 
between these equations and thermodynamics. 
}

\blue{
Based on Sekimoto's ideas, Seifert \cite{seifert2005entropy} established the 
fluctuation theorem in 2005, which provides a detailed description of entropy 
production fluctuations in small systems. Shortly thereafter, Seifert introduced 
the term \emph{stochastic thermodynamics} \cite{seifert2008stochastic} for the 
thermodynamics based on stochastic process, which is now more widely used. 
Stochastic thermodynamics has since become a prominent area within non-equilibrium 
thermodynamics. Although this field has been developing for nearly thirty years, 
there are still many interesting issues which are worth of further 
exploration\cite{peliti2021stochastic,shiraishi2023introduction}.
}

Apart from the current hot topics such as discussing specific bounds based on 
this theory \cite{dechant2021improving,koyuk2021quality}, a more fundamental 
question is whether we can discuss this theory within the framework of relativity. 
It is well known that all physics laws should inherently abide by the principles of 
relativity. However, the original Langevin equation was built on 
top of Newtonian mechanics. This situation calls for a relativistic covariant treatment 
for the stochastic motion of the Brownian particle. Recently, we 
established a reasonable framework for relativistic stochastic mechanics
in \cite{cai2023relativisticI,cai2023relativisticII}, in which both the Langevin  
and Fokker-Planck equations (FPEs) are fully general relativistic covariant and 
are independent on the concrete choice of spacetime background except for a 
requirement of stationarity. As a natural continuation of these preceding works, 
it seems now the right time to revisit the existing stochastic thermodynamics within 
the framework of general relativity. 

The related problem in the context of special relativity has been discussed by Pal 
and Deffner \cite{pal2020stochastic}. Our work extends the relevant discussions to 
curved spacetime. Some unnatural assumptions are discarded in our analysis. 
The clarification for the connection and difference between the coordinate system 
and the observer in \cite{cai2023relativisticI} provides us a solid working ground 
to maintain covariance throughout the formalism. Thanks to such generality, 
our discussion is irrelevant to any specific choice of spacetime background. 

This paper is organized as follows. We review the covariant formalism of 
relativistic stochastic mechanics in Section \ref{sec2}, and the corresponding content 
has been extended to the case of charged Brownian particle under the influence of 
external Maxwell field. In Section \ref{sec3} and Section \ref{sec4}, we discuss the 
first and second laws of stochastic thermodynamics. To maintain fully 
relativistic covariance, these laws are presented in the form of divergences of 
the energy and entropy density currents, and the powers of gravity and of the 
electromagnetic field as well as the heat transfer rate play essential roles 
in the first law. Section \ref{sec5} is devoted to the analysis on the stability 
of the equilibrium state, which helps to elucidate the authentic underpinnings 
of the long time limitation. In Section \ref{sec6}, we present a brief 
summary and outline some open questions that still need to be answered along the line 
of the current research.

The convention of notations follows our previous work 
\cite{cai2023relativisticI,cai2023relativisticII}, and it is important 
to distinguish random and deterministic variables, as well as objects in 
different spaces. The variables with tilde are random variables, and the 
corresponding un-tilded symbols represent their concrete realizations. 
The objects in different spaces (including the $(d+1)$-dimensional spacetime
$\mathcal{M}$, future mass shell bundle $\Gamma_{m}^{+}$ and the future mass 
shell $(\Gamma_{m}^{+})_{x}$ at the event $x$) are distinguished by the font of the objects and 
their indices. We will provide the specific meaning of new symbols when they appear.

\section{Elements of relativistic stochastic mechanics}
\label{sec2}

\blue{
The characteristic equation of motion in stochastic mechanics is the 
Langevin equation which describes the motion of a heavy particle 
under the influence of a heat reservoir, specifically through the 
interaction of a damping force and a stochastic force. In the general relativistic 
context, we have elucidated that the damping and stochastic forces should 
take the following form \cite{cai2023relativisticI}:
\begin{align}
\mathcal{F}_{\text{damp}}^\mu = \mathcal{K}^{\mu\nu}U_\nu, \qquad 
\mathcal{F}_{\text{stochastic}}^\mu = \mathcal{F}^{\mu}_{\text{add}} 
+ \mathcal{R}^{\mu}{}_{\mathfrak{a}} \circ_{S} \D \tilde{w}_{\tau}^\mathfrak{a}/{\D\tau},
\end{align}
where $\mathcal{R}^{\mu}{}_{\mathfrak{a}}$, $\mathcal{F}^{\mu}_{\text{add}}$, 
and $\mathcal{K}^{\mu\nu}$ respectively denote the amplitudes of the 
Gaussian stochastic noise $\D\tilde{w}_\tau^\mathfrak{a}$, 
the additional stochastic force, and the friction tensor. To maintain the chain rule 
under general coordinate transformation, we adopted the Stratonovich coupling $\circ_S$
between the stochastic amplitude and the Gaussian noise. 
Supplementing these forces into the geodesic equation yields the covariant 
Langevin equation LE$_{\tau}$ which takes the proper time $\tau$ of the 
Brownian particle as evolution parameter,}
\begin{align}
    \left\{
    \begin{aligned}
    \D \tilde{x}_{\tau}^{\mu} &= \frac{\tilde{p}_{\tau}^{\mu}}{m}\D \tau, \\
    \D \tilde{p}_{\tau}^{\mu} &= \left[\mathcal{R}^{\mu}{}_{\mathfrak{a}} \circ_{S} \D \tilde{w}_{\tau}^\mathfrak{a} + \mathcal{F}^{\mu}_{\text{add}}\D \tau\right] + \mathcal{K}^{\mu\nu}U_{\nu}\D \tau - \frac{1}{m}\varGamma^{\mu}{}_{\alpha\beta}\tilde{p}_{\tau}^{\alpha}\tilde{p}_{\tau}^{\beta}\D \tau.
    \end{aligned}
    \right.
\end{align}
Here, $(\tilde{x}^{\mu}_{\tau},\tilde{p}^{\mu}_{\tau})$ represents the path of the 
Brownian particle in the tangent bundle $T\mathcal{M}$, or more precisely in the 
future mass shell bundle $\Gamma_m^+ =\bigcup_{x\in {\mathcal M}} (\Gamma_m^+)_x$. Thanks to the mass shell condition, 
the components of $\tilde{p}^{\mu}_{\tau}$ are not all independent. 
The assumption that the stationary solution of the corresponding FPE is 
identical to the one particle distribution function (1PDF) of the reservoir particles 
in the long time limit helps in determining the form of the additional 
stochastic force \cite{cai2023relativisticII} 
\begin{align}\label{Fadd}
\mathcal{F}^{\mu}_{\text{add}}
=\frac{\delta^{\fa\fb}}{2}\mathcal{R}^{\mu}{}_{ \mathfrak{a}}
\nabla^{(h)}_{j}\mathcal{R}^{j}{}_{ \mathfrak{b}},
\end{align}
where $\nabla^{(h)}_{j}$ is the covariant derivative on $(\Gamma_{m}^{+})_{x}$. 
$U_{\nu}$ is the proper velocity of the heat 
reservoir and also of the reservoir comoving observer {\em Bob}. 
$\varGamma^{\mu}{}_{\alpha\beta}$ appearing in the last term is the Christoffel 
connection on $\mathcal{M}$. Notice that, in eq.\eqref{Fadd}, 
$\mathcal{R}^{\mu}{}_{ \mathfrak{a}}$ and $\mathcal{R}^{i}{}_{ \mathfrak{a}}$ \vspace{3pt}
are the same thing but with different coordinate bases 
$\displaystyle \left\{\pfrac{}{p^{\mu}}\right\}$ and \vspace{3pt}
$\displaystyle \left\{\pfrac{}{\breve{p}^{i}}
:=\pfrac{}{{p}^{i}} -\dfrac{p_{i}}{p_{0}}\pfrac{}{{p}^{0}}
\right\}$, 
and all tensors on the mass shell $(\Gamma_{m}^{+})_{x}$ have such 
different representations.

In general, the proper time $\tau$ of the Brownian particle should become a random variable 
with respect to an arbitrarily chosen, non-comoving observer {\em Alice}. Therefore,
a better description for the stochastic motion of the Brownian particle 
should be parametrized by the proper time $t$ of Alice, rather than that of 
the Brownian particle. This 
is achieved in \cite{cai2023relativisticI,dunkel2009time} through a reparametrization scheme and the 
corresponding Langevin equation is denoted LE$_t$. The concrete form of LE$_t$
will not be used in this work, so we omit it.

In order to establish a statistical description for the Brownian particle, the first 
thing to be clarified is the space of states. It is crucial to notice that 
the configuration space must be taken as an ``equal time slice'' of the spacetime 
manifold, and as such it is inherently observer dependent. To keep things as 
general as possible, we adopt Alice as the observer whose proper velocity is 
denoted as  $Z^{\mu}$. The proper time $t$ of Alice is made use of in making the 
time slicing,
\begin{align}
\mathcal{S}_{t}=\{x\in {\mathcal M}| t(x)=t={\rm const}.\},
\end{align}
where $t(x)$ is an extension of the proper time $t$ over $\mathcal{M}$ as a 
scalar field. The space of states is a hypersurface $\Sigma_{t}$ in the 
mass shell bundle \cite{cai2023relativisticII},
\begin{align}
\Sigma_{t}:=\bigcup_{x\in\mathcal{S}_{t}} (\Gamma^{+}_{m})_{x}
=\{(x,p)\in\Gamma^{+}_{m}|x\in\mathcal{S}_{t}\},
\end{align}
which is accompanied by the timelike hypersurface orthogonal vector field 
$\mathscr{Z}=Z^{\mu}e_{\mu}$, where $e_{\mu}$ is a part of the orthogonal basis of 
the mass shell bundle $\{e_{\mu},\partial/\partial p^{\mu}\}$. 
As mentioned earlier, LE$_\tau$ is not the proper version of the equation 
that describes the stochastic motion in $\Sigma_{t}$ and we should replace it by LE$_t$. 
From the point of view of stochastic thermodynamics, the more important object is the corresponding 
FPE. After conducting both numerical and analytical validations 
\cite{cai2023relativisticI,cai2023relativisticII}, it has been proved that the physical 
distribution $\varphi$ of the Brownian particle as observed by Alice 
is encoded in the solution of the reduced FPE,
\begin{align}\label{reduced-fp}
\frac{1}{m}\mathscr{L}(\varphi)=\nabla^{(h)}_{i}\mathcal I^{i}[\varphi],
\end{align}
where 
\[
\mathscr{L}=p^{\mu}\dfrac{\partial}{\partial x^{\mu}}
-\varGamma^{\mu}{}_{\alpha\beta}p^{\alpha}p^{\beta}\dfrac{\partial}{\partial p^{\mu}}
\]
is the Liouville vector field which is also the Hamiltonian vector field 
associated with the Hamiltonian $\displaystyle H=\frac{1}{2m} g_{\mu\nu} p^\mu p^\nu$ 
of a free relativistic particle,
\blue{
\begin{align}\label{heat-crrent}
\mathcal{I}[\varphi]
:=\left[\frac{1}{2}\mathcal{D}^{ij}\nabla^{(h)}_{j}\varphi
-\mathcal{K}^{i\nu}U_{\nu}\varphi\right]\pfrac{}{\breve{p}^i}
=\left[\frac{1}{2}\mathcal{D}^{\mu j}\nabla^{(h)}_{j}\varphi
-\mathcal{K}^{\mu\nu}U_{\nu}\varphi\right]\pfrac{}{p^\mu},
\end{align}
}and $\mathcal{D}^{ij}=\delta^{\mathfrak{ab}}
\mathcal{R}^{i}{}_{ \mathfrak{a}}\mathcal{R}^{j}{}_{ \mathfrak{b}}$ 
is the diffusion tensor in the momentum space. 
\blue{A notation convention is worth mentioning: $\mathscr{L}(\varphi)$ represents 
the Liouville vector field acting on $\varphi$ as a directional derivative, 
while $\mathcal{I}[\varphi]$ denotes a vector dependent on $\varphi$. 
The vector field $\mathcal{I}[\varphi]$ is tangent to the mass shell, thus its 
components can carry either a Latin index or a Greek one, corresponding to different bases.}
Moreover, this vector field is closely connected to the heat transfer rate 
from the reservoir to the Brownian particle,
\begin{align}
Q[\varphi]:=\int_{(\Gamma^+_m)_x}\eta_{(\Gamma^+_m)_x} Z_\nu \mathcal I^\nu[\varphi].
\label{heat-current}
\end{align}
When $\mathcal{I}^{i}[\varphi]$ vanishes, the heat transfer rate also vanishes. 
\blue{More details about the reduced Fokker-Planck equation can be found in 
\cite{cai2023relativisticII}.}

Some characteristic quantities of the system including particle number 
current \blue{$N^\mu[\varphi]$} and energy-momentum-stress tensor 
\blue{$T^{\mu\nu}[\varphi]$} can be constructed using this distribution,
\begin{align}\label{n-T}
N^{\mu}[\varphi]=\int\eta_{(\Gamma_{m}^{+})_{x}}\dfrac{p^{\mu}}{m}\varphi N
:=Nn^{\mu}[\varphi],
\qquad 
T^{\mu\nu}[\varphi]=\int\eta_{(\Gamma_{m}^{+})_{x}}\dfrac{p^{\mu}p^{\nu}}{m}\varphi,
\end{align}
where $N$ is a constant denoting the total number of the Brownian particle and $n^{\mu}$ is 
the one-particle current. Here and below we use the notation $\eta_{(\Gamma_{m}^{+})_{x}}$ 
to denote the invariant volume element on $(\Gamma_{m}^{+})_{x}$. Similar notations 
such as $\eta_{\mathcal M}$, $\eta_{{\mathcal S}_t}$, $\eta_{\Sigma_t}$ {\em etc} 
are all invariant volume elements on the relevant manifolds given in the suffices.
The divergences of the above tensors are given as follows \cite{cai2023relativisticII},
\begin{align}
\nabla_{\mu}N^{\mu}[\varphi]=N\nabla_{\mu}n^{\mu}[\varphi]=0,
\qquad 
\nabla_{\mu}T^{\mu\nu}[\varphi]=-\int\eta_{(\Gamma_{m}^{+})_{x}}\mathcal{I}^{\nu}[\varphi].
\end{align}

In the rest of this section, we will extend the basic equations of 
stochastic mechanics in curved spacetime to the case of 
charged Brownian particle subjecting to external 
electromagnetic field $A_\mu$. Let $F_{\mu\nu}= \partial_\mu A_\nu-\partial_\nu A_\mu$ 
be the corresponding field strength, then 
the extension of LE$_\tau$ should read
\begin{align}\label{LEtau-em}
\left\{\begin{aligned}
\D \tilde{x}_{\tau}^{\mu}&=\frac{\tilde{p}_{\tau}^{\mu}}{m}\D \tau,\\
\D \tilde{p}_{\tau}^{\mu}
&=\left[\mathcal{R}^{\mu}{}_{ \mathfrak{a}}\circ_{S}\D \tilde{w}_{\tau}^\mathfrak{a}
+\mathcal{F}^{\mu}_{\text{add}}\D \tau\right]
+\mathcal{K}^{\mu\nu}U_{\nu}\D \tau
-\frac{1}{m}\varGamma^{\mu}{}_{\alpha\beta}\tilde{p}_{\tau}^{\alpha}\tilde{p}_{\tau}^{\beta}\D \tau
+\frac{q}{m}F^{\mu}{}_{\nu}\tilde{p}^{\nu}_{\tau}\D\tau.
\end{aligned}\right.
\end{align}
The modification to the reduced FPE is fully encoded in the modified Liouville vector field
\cite{sarbach2014tangent}
\begin{align}
\mathscr{L}_{F}=p^{\mu}\dfrac{\partial}{\partial x^{\mu}}
+\(qF^{\mu}{}_{\nu}p^{\nu}-\varGamma^{\mu}{}_{\alpha\beta}
p^{\alpha}p^{\beta}\)\dfrac{\partial}{\partial p^{\mu}}.
\end{align}
Apart from the different form of the Liouville vector field, the reduced FPE 
takes the same form as eq.\eqref{reduced-fp}, which can be verified by use of the 
diffusion operator approach (see the appendix in \cite{cai2023relativisticII}),
\begin{align}\label{reduced-fp-em}
\dfrac{1}{m}\mathscr{L}_{F}(\varphi)=\nabla_{i}^{(h)}\mathcal{I}^{i}[\varphi].
\end{align}
In particular, the vector field $\mathcal{I}^{i}[\varphi]$ maintains 
its original form as given in eq.\eqref{heat-crrent}. 
Using the probability current of the Brownian particle defined as
\begin{align}\label{pro-current}
  \mathscr{J}[\varphi]:=\frac{\varphi}{m}\mathscr{L}_F-\mathcal{I}[\varphi],
\end{align}
the reduced FPE \eqref{reduced-fp-em} can also be written as a 
current conservation equation
\begin{align}
\hat\nabla^{(\hat h)}_A\mathscr{J}^A[\varphi]=0,
\end{align}
where $\hat\nabla^{(\hat h)}$ is covariant derivative operator on the mass shell bundle.

We assume that the heat reservoir is in its intrinsic equilibrium state with the 1PDF 
\cite{degroot1980relativistic,sarbach2014tangent,acuna2022introduction,
cercignani2002relativistic}
\[
\varphi_R = \re^{-\alpha_R +U_\mu p^\mu/T_B},
\]
where \blue{$T_B$ is the temperature as measured by the comoving observer Bob, 
and $\alpha_{R}=\mu_B/T_B$ is the ratio between the chemical potential and 
temperature of the reservoir particles. Unlike flat spacetime, 
the temperature $T_B$ and chemical potential $\mu_B$ of a thermal equilibrium 
system in curved spacetime are not constants. Instead, the equilibrium condition 
of a gravitational system should become the generalized gradient of the temperature 
and the chemical potential is zero \cite{liu2022covariant,hao2024Gravito}:
\begin{align}
  \mathcal{D}_\mu T_B=\nabla_\mu T_B+T_B U^\nu\nabla_\nu U_\mu=0,\qquad
  \mathcal{D}_\mu \mu_B=\nabla_\mu \mu_B+\mu_B U^\nu\nabla_\nu U_\mu=0.
\end{align}
Therefore, both $T_B$ and $\alpha_R$ should be viewed as scalar fields 
on the spacetime manifold.}  
In the long time limit, the 1PDF for the Brownian particle should approach the same 
form as the equilibrium distribution of the reservoir, 
i.e. 
\begin{align}\label{eqdist}
\varphi_{\text{eq}}=\e^{-\alpha+\beta_{\mu}p^{\mu}}, \qquad
\beta_\mu = U_\mu/T_B,
\end{align}
and it should stop receiving heat from the reservoir, i.e. 
$\mathcal{I}^{i}[\varphi]=0$. \blue{It should be noted that due to the heat exchange 
between the Brownian particle and the thermal reservoir, the thermal equilibrium 
state of the Brownian particle has the same temperature as that of the reservoir. 
However, since there is no chemical reaction between the Brownian particle and the 
reservoir, their $\alpha$ do not need to be the same.}
On such occasions, the reduced FPE reduces into the Liouville equation 
$\mathscr{L}_{F}(\varphi)=0$, which leads to the following constraints over
$\alpha$ and $\beta_\mu$ (the equilibrium conditions)
\begin{align}
	\nabla_{\mu}\alpha+q\beta^{\nu}F_{\mu\nu}=0,\qquad
	\nabla_{(\mu}\beta_{\nu)}=0.\label{killing}
\end{align}
The condition over $\beta_{\mu}$ remains the same as in the case of neutral Brownian particle,
which implies that $\beta_{\mu}$ should be future directed and timelike Killing. 
The condition over $\alpha$ is dependent on the electromagnetic field strength, thus 
making a difference from the neutral case. 

It is easy to check that the condition $\mathcal{I}^{i}[\varphi]=0$ 
calls for an un-altered Einstein relation as in the case of 
neutral Brownian particle,
\begin{align}\label{einstein-relation}
	\mathcal{D}^{ij}=2T_{B}\mathcal{K}^{ij},
\end{align}
and the particle number current remains divergence-free. However, 
the divergence of the energy-momentum tensor needs to be modified 
which reflects the consequence of the action of the electromagnetic field,
\begin{align}\label{divergence-T-em}
	\nabla_{\mu}T^{\mu\nu}[\varphi]=q F^{\nu}_{\ \mu}n^{\mu}[\varphi]-\int\eta_{(\Gamma_{m}^{+})_{x}}\mathcal{I}^{\nu}[\varphi].
\end{align}

\section{First law of relativistic stochastic thermodynamics}
\label{sec3}

The first law of relativistic Brownian particle in the absence of an 
external field has been established in our previous work \cite{cai2023relativisticII}, 
which is presented in the form of the divergence of the energy current rather 
than in the change in the energy itself as in the ordinary thermodynamics,
\begin{align}\label{noex-firstlaw}
\nabla_{\mu} E^{\mu}[\varphi]=P_{\text{grav}}[\varphi]+Q[\varphi],
\end{align}
where $Q[\varphi]$ is the heat transfer rate as given in eq.\eqref{heat-current}, and
\begin{align}
P_{\text{grav}}[\varphi]
&=-\int\eta_{(\Gamma_{m}^{+})_{x}} \dfrac{\varphi}{m}p^{\mu}p^{\nu}\nabla_{\mu}Z_{\nu}
=-T^{\mu\nu}[\varphi]\nabla_\mu Z_\nu
\label{Pgrav}
\end{align}
is the gravitational power as measured by Alice \cite{liu2020work}. 
In this section, we will build the complete first law of relativistic Brownian particle 
in the presence of electromagnetic field. 

It should be mentioned that, in the presence of electromagnetic field, the definition of
energy of the charged Brownian particle is non-unique. Such non-uniqueness 
arises from different choices of the proper momentum of the particle, i.e. 
the kinematic momentum $p^\mu$ (which is proportional to the proper velocity of the particle)
and the physical momentum $P^\mu= p^{\mu}+q A^{\mu}$. Each choice gives rise to a 
separate definition of the energy, i.e.
\begin{align}
	E_{p}=-Z_{\mu}p^{\mu}
\end{align}
and \cite{landau1975classical}
\begin{align}
	E_{P}=-Z_{\mu}P^{\mu}.
\end{align}
$E_{p}$ corresponds to the kinematic energy and $E_P$ corresponds to the Hamiltonian. 
It will be clear that the different choices of energy leads to different forms of the 
first law of relativistic stochastic thermodynamics, 
both forms are physically correct. 

Following the discussions made in Appendix~\ref{app.A}, we now introduce the energy currents 
associated with $E_{p}$ and $E_{P}$ as follows,
\begin{align}\label{energy-current}
E_{p}^{\mu}[\varphi]=\int\eta_{(\Gamma_{m}^{+})_{x}}\dfrac{p^{\mu}}{m}\varphi E_{p},\qquad
E_{P}^{\mu}[\varphi]=\int\eta_{(\Gamma_{m}^{+})_{x}}\dfrac{p^{\mu}}{m}\varphi E_{P}.
\end{align} 
These currents are related to the energy-momentum tensor via the following equations,
\begin{align}
E_{p}^{\mu}[\varphi]=-Z_{\nu}T^{\mu\nu}[\varphi],\qquad
E_{P}^{\mu}[\varphi]=-Z_{\nu}(T^{\mu\nu}[\varphi]+qA^{\mu}n^{\nu}[\varphi]),
\end{align} 
where $T^{\mu\nu}[\varphi]$ and $n^{\mu}[\varphi]$ are given in eq.\eqref{n-T}. 
Using the trick discussed in Appendix~\ref{app.A}, the divergences of the above currents
are calculated to be
\begin{align}\label{firstlaw-p}
\nabla_{\mu}E_{p}^{\mu}[\varphi]
&=-\int\eta_{(\Gamma_{m}^{+})_{x}} 
\(\dfrac{\varphi}{m}p^{\mu}p^{\nu}\nabla_{\mu}Z_{\nu}
+\dfrac{q\,\varphi}{m}F_{\mu\nu}p^{\nu}Z^{\mu}-\mathcal{I}^{\mu}[\varphi]Z_{\mu}\),
\end{align} 
and
\begin{align}\label{firstlaw-P}
\nabla_{\mu}E_{P}^{\mu}[\varphi]
&=-\int\eta_{(\Gamma_{m}^{+})_{x}} \(\dfrac{\varphi}{m}p^{\mu}p^{\nu}\nabla_{\mu}Z_{\nu}
+\dfrac{q\,\varphi}{m}p^{\mu}\pounds_{Z}A_{\mu}-\mathcal{I}^{\mu}[\varphi]Z_{\mu}\),
\end{align} 
in which $\pounds_{Z}A_{\mu}$ is the Lie derivative of $A_\mu$ along 
the direction of $Z^\mu$. The first and the third terms on the right hand side 
of eqs.\eqref{firstlaw-p} and \eqref{firstlaw-P} can be easily recognized to 
be the gravitational power and the heat transfer rate respectively
as presented in eqs.\eqref{Pgrav} and \eqref{heat-crrent}. The middle terms 
on the right hand side of eqs.\eqref{firstlaw-p} and \eqref{firstlaw-P} 
can be respectively denoted as
\begin{align}
P_{\text{em}}[\varphi]
&=-\int\eta_{(\Gamma_{m}^{+})_{x}} \dfrac{q\,\varphi}{m}F_{\mu\nu}p^{\nu}Z^{\mu}
=-qF_{\mu\nu}n^\mu[\varphi]Z^\nu ,\nonumber\\
P_{\text{nc}}[\varphi]
&=-\int\eta_{(\Gamma_{m}^{+})_{x}} \dfrac{q\,\varphi}{m}p^{\mu}\pounds_{Z}A_{\mu}
=-qn^\mu[\varphi]\pounds_{Z}A_\mu.
\end{align}
$P_{\text{em}}[\varphi]$ is obviously the power of the electromagnetic field, 
while $P_{\text{nc}}[\varphi]$ should be understood as the power of the
non-conservative part of the electromagnetic 
field observed by Alice. It might be helpful to point out that 
$P_{\text{nc}}[\varphi]\neq 0$ only when $\pounds_{Z}A_\mu\neq 0$, i.e. 
when the proper velocity of Alice is not a symmetry generator of the electromagnetic field.
The difference between eq.\eqref{firstlaw-p} and eq.\eqref{firstlaw-P} lies in the 
fact that $E_{P}$ absorbs the potential energy of the particle into the total energy. 

Finally, both versions of the first law can be expressed in a concise form,
\begin{align}
&\nabla_{\mu}E_{p}^{\mu}=P_{\text{grav}}[\varphi]+P_{\text{em}}[\varphi]+Q[\varphi],\\
&\nabla_{\mu}E_{P}^{\mu}=P_{\text{grav}}[\varphi]+P_{\text{nc}}[\varphi]+Q[\varphi].
\end{align}

\section{Second law of relativistic stochastic thermodynamics}
\label{sec4}

The research on macroscopic irreversibility has taken central stage in the history of 
thermodynamics and statistical physics. After about 150 years of development, 
this field has grown considerably. For example, it was found that the second law of 
thermodynamics can also be established in stochastic thermodynamics 
\cite{sekimoto1998langevin}, and we will give a relativistic version of the law in 
this section. It is worth mentioning that the fluctuation theorem gives a fairly good 
explanation for the origin of macroscopic irreversibility 
\cite{kurchan1998fluctuation,searles1999fluctuation,crooks1999entropy,hatano2001steady,chernyak2006path,seifert2012stochastic}, 
but the relevant analysis is relatively complex, so we leave it to a 
separate upcoming work. 

Let us now precede by separating the probability current \eqref{pro-current} 
of the Brownian particle into the time-reversible and dissipative parts 
as did in \cite{risken1996fokker,esposito2010three,van2010three},
\begin{align}\label{r-d-current}
\mathscr{J}_{\text{r}}[\varphi]=\dfrac{\varphi}{m}\mathscr{L}_{F},\qquad
\mathscr{J}_{\text{d}}[\varphi]=-\mathcal{I}[\varphi].
\end{align}
The reversibility of $\mathscr{J}_{\text{r}}[\varphi]$ can be easily understood because 
$\mathscr{L}_{F}$ is the Hamiltonian vector field which is clearly time-reversible. 
The dissipative part is caused by the diffusion and friction forces which 
breaks the time-reversal symmetry.
It will be clear shortly that the entropy 
production is closely related to the breaking of this symmetry.

The definition of entropy in non-equilibrium states is not as clear as the entropy 
in the equilibrium \cite{goldstein2019nonequilibrium}, especially in the 
relativistic context. We could of course imitate the form of Gibbs entropy 
but since we have at least three different FPEs for different 
distribution functions \cite{cai2023relativisticII}, and each distributions 
have reasonable meanings in different spaces, it seems that the definition 
for the entropy is not unique. Fortunately, to fulfill our aim to 
formulate the second law of relativistic stochastic thermodynamics, 
it suffices to have one definition for the entropy which is non-decreasing 
in the course of time. Following the discussions made in \cite{cai2023relativisticII},
it is reasonable to start with the 1PDF $\varphi$ which obeys the reduced FPE. 
This object is the physical distribution on the future mass shell bundle. 
Therefore, we introduce the entropy density on the 
mass shell bundle of the Brownian particle as
\begin{align}
s(x,p)=-\log\varphi(x,p).
\end{align}
This entropy density characterizes the 
properties of the micro states of the Brownian particle. 
In stochastic thermodynamics, the entropy production comes not only from 
the Brownian particle but also from the heat reservoir. 
Therefore, the relative entropy is more suitable 
to portray the irreversibility of the whole system. The  
relative entropy density reads 
\begin{align}
h(x,p)=-\log\dfrac{\varphi(x,p)}{\varphi_{\text{eq}}(x,p)}
=-\log\varphi-\alpha+\beta_{\mu}p^{\mu},
\end{align} 
where $\varphi_{\text{eq}}$ is the equilibrium distribution of the Brownian particle, 
hence $\alpha$ and $\beta_{\nu}$ satisfy the equilibrium condition \eqref{killing}. 

What we actually need to make use of while formulating the second law of 
relativistic stochastic thermodynamics is not the the total entropy but rather the
entropy density current on the space time. 
The entropy density current of the Brownian particle on the spacetime is given as
\begin{align}\label{entropy-current}
S^{\mu}[\varphi]=-\int \eta_{\(\Gamma_{m}^{+}\)_{x}} \dfrac{p^{\mu}}{m} \varphi \log \varphi,
\end{align}
and the relative entropy density current can be expressed as the linear combination 
of entropy density current, the particle number current, and the energy-momentum tensor 
\begin{align}
	H^{\mu}[\varphi]=S^{\mu}[\varphi]-\alpha n^{\mu}[\varphi]+\beta_{\nu}T^{\mu\nu}[\varphi].
\end{align}
We will shortly prove that the relative entropy production rate $\nabla_\mu H^\mu[\varphi]$ 
is the sum of the entropy production rate of the Brownian particle and 
that of the heat reservoir.

Since the heat reservoir is assumed to be in local thermodynamic equilibrium, 
its entropy current can be expressed as \cite{cercignani2002relativistic,hao2022relativistic} 
\begin{align}\label{SNT-reservoir}
S^{\mu}_{R}=(1+\alpha_{R})N^{\mu}_{R}-\beta_{\nu} T^{\mu\nu}_{R},
\end{align}
\blue{where $S^{\mu}_{R}$, $N^{\mu}_{R}$ and $T^{\mu\nu}_{R}$ are respectively
the entropy current density, particle numbers current density and 
energy-momentum current density for the reservoir particles.}
This expression is not affected by the presence of the electromagnetic field, but the 
equilibrium condition of the reservoir will take a form similar to that of the 
Brownian particle, 
\begin{align}\label{equilibrium-condition-r}
\nabla_{\mu}\alpha_{R}+q_{R}\beta^{\nu}F_{\mu\nu}=0,\qquad
\nabla_{(\mu}\beta_{\nu)}=0,
\end{align}  
where $q_{R}$ characterizes the charge of the reservoir particle 
which could either be vanishing or not. We assume that the reservoir 
is consisted purely of classical particles with no chemical reactions, 
its particle number current should be conserved, $\nabla_\mu N^\mu_R=0$. However, 
since the reservoir exchanges energy with the Brownian particle and also with 
the electromagnetic field, the divergence of the total energy-momentum tensor needs to obey
\begin{align}\label{energy-conservation}
\nabla_{\mu} T^{\mu\nu}_{R}+\nabla_{\mu} T^{\mu\nu}[\varphi]
=F^{\nu}{}_{\mu}(q n^{\mu}[\varphi]+q_{R}N_{R}^{\mu}),
\end{align}
which means that the non-conservation of the total energy-momentum tensor is originated solely from 
the external field. With the help of eqs.(\ref{SNT-reservoir}--\ref{energy-conservation}) 
and eq.\eqref{divergence-T-em}, we finally find the covariant divergence of entropy density 
current of the reservoir,
\begin{align}\label{divergence-Sr}
\nabla_{\mu}S_{R}^{\mu}&=N_{R}^{\mu}\nabla_{\mu}\alpha_{R}
-\beta_{\nu}\nabla_{\mu}T_{R}^{\mu\nu}\nonumber\\
&=-q_{R}N_{R}^{\mu}\beta^{\nu}F_{\mu\nu}-\beta_{\nu}
\left[F^{\nu}{}_{\mu}(q n^{\mu}+q_{R}N_{R}^{\mu})
-\nabla_{\mu}T^{\mu\nu}[\varphi]\right]\nonumber\\
&=-n^{\mu}[\varphi]\nabla_{\mu}\alpha+\beta_{\nu}\nabla_{\mu}T^{\mu\nu}[\varphi],	
\end{align}
where the equilibrium condition of reservoir particles \eqref{equilibrium-condition-r} 
is used in the second line and the equilibrium condition of Brownian particle \eqref{killing} 
is used in the third line. The total entropy production rate is then
\begin{align}
\nabla_{\mu}S^{\mu}_{\text{tot}}  
&=\nabla_{\mu}S^{\mu}[\varphi]+\nabla_{\mu}S_{\text{R}}^{\mu}\nonumber\\
&=\nabla_{\mu}S^{\mu}[\varphi]-n^{\mu}[\varphi]\nabla_{\mu}\alpha
+\beta_{\nu}\nabla_{\mu}T^{\mu\nu}[\varphi]=\nabla_{\mu}H^{\mu}[\varphi],
\end{align}
which is the desired result advocated earlier.

Using eq.\eqref{divergence} in Appendix~\ref{app.A} we can evaluate the 
total entropy production rate to be
\begin{align}\label{secondlaw-h}
\nabla_\mu S^\mu_{\text{tot}}=\nabla_{\mu}H^{\mu}[\varphi]
&=\int\eta_{(\Gamma^{+}_{m})_{x}} \mathscr{J}[\varphi](h)\nonumber\\
&=\int\eta_{(\Gamma_{m}^{+})_{x}}\dfrac{\mathcal{D}^{\mu\nu}}{2\varphi}
\(\pfrac{}{p^\mu}\varphi-\beta_{\mu}\varphi\)\(\pfrac{}{p^\nu}\varphi-\beta_{\nu}\varphi\),
\end{align}
where the Einstein relation and the vanishing boundary condition for $\varphi$ at 
the infinity of the mass shell have been used. Since the diffusion tensor $\mathcal{D}^{ij}
=\delta^{\mathfrak{ab}}\mathcal{R}^{i}{}_{\mathfrak{a}}\mathcal{R}^{j}{}_{\mathfrak{b}}$ 
must be a semi-positive definite quadratic form, the total entropy production rate 
is always nonnegative. Therefore, we arrive at the second law of 
relativistic stochastic thermodynamics
\begin{align}
  \nabla_\mu S^\mu_{\text{tot}}\ge 0.
\end{align}

The entropy production is always closely related to the breaking of the 
time-reversal symmetry. We can divide the total entropy production rate into that of the 
Brownian particles and of the heat reservoir, both will be shown to be proportional to 
the dissipative part of the probability current.

The entropy production rate of the Brownian particle reads
\begin{align}\label{entropy-production}
\nabla_{\mu}S^{\mu}[\varphi]
&=\int\eta_{(\Gamma_{m}^{+})_{x}}\mathscr{J}[\varphi](-\log\varphi)\nonumber\\
&=-\int\eta_{(\Gamma_{m}^{+})_{x}}
\varphi^{-1}\(\pfrac{\varphi}{p^\mu}\)\mathscr{J}^\mu_{\mathrm{d}}[\varphi],
\end{align}
where the trick mentioned in Appendix~\ref{app.A} is used again, 
and this derivation is similar to that leads to eq.\eqref{secondlaw-h}. 
Substituting eq.\eqref{divergence-T-em} into eq.\eqref{divergence-Sr}, 
the entropy production rate of the reservoir caused by the heat exchange
can be expressed as 
\begin{align}\label{entropyproductionofheat}
\nabla_{\mu}S_{R}^{\mu}&=-\beta_{\nu}\int\eta_{(\Gamma_{m}^{+})_{x}}\mathcal{I}^{\nu}
[\varphi]=\int\eta_{(\Gamma_{m}^{+})_{x}}\beta_{\mu}\mathscr{J}^{\mu}_{\text{d}}[\varphi].
\end{align}
Similar result was also mentioned in Pal and Deffner's paper \cite{pal2020stochastic}, 
but now it is made fully covariant. 
It is obvious that the time-reversible part $\mathscr{J}_{\mathrm{r}}^\mu[\varphi]$ 
of the probability current does not contribute to the entropy production rate. 
All entropy productions arise as consequences 
of the dissipative part $\mathscr{J}_{\D}^\mu[\varphi]$, which is the same as 
in the non-relatrivistic case \cite{van2010three,pal2020stochastic}. 
It is worth noticing that eq.\eqref{entropyproductionofheat} is actually 
the relativistic version of Clausius' identity for the heat reservoir,
\begin{align}
\nabla_\mu S^\mu_{R}=-\frac{Q_B[\varphi]}{T_B}=\frac{Q_R}{T_B},
\end{align}
where $Q_B[\varphi]$ is the heat transfer rate from the heat reservoir to the Brownian 
particle from the perspective of Bob, and hence $Q_R=-Q_B[\varphi]$ 
is the heat transfer rate from the Brownian particle to the heat reservoir. 

There are some interesting consequences from the equation of entropy production rate. 
Assuming that the diffusion tensor $\mathcal{D}^{ij}$ is a full-rank matrix 
and hence strictly positive definite, the condition for 
zero entropy production rate becomes
\begin{align}
\(\pfrac{}{p^\mu}-\beta_{\mu}\)\varphi=0.
\end{align}
This is equivalent to the thermal equilibrium condition $\mathcal{I}[\varphi]=0$,
thanks to the covariant Einstein relation \eqref{einstein-relation}. 
Therefore, the equilibrium state is the unique state which saturates the second law. 
However, if the diffusion tensor is not full-rank, implying that there exists 
some nonzero vector field $\mathcal{C}^i[\varphi]$ on the momentum space obeying 
$\mathcal{D}_{ij}\mathcal{C}^i[\varphi]\mathcal{C}^j[\varphi]=0$, the condition for 
zero entropy production rate will become 
\begin{align}
\(\pfrac{}{p^\mu}-\beta_{\mu}\)\varphi=\mathcal{C}_{\mu}[\varphi].
\end{align}
This leads to the possibility for the existence of certain state which is 
different from the prescribed equilibrium state but still with zero entropy production rate. 
This situation is not unexpected. The presence of a degenerate diffusion tensor 
implies the existence of certain degrees of freedom that are decoupled from  
the heat reservoir. The 1PDF of these states obeys the Liouville equation and  
the entropy remains conserved. In the next section, we will discuss how 
the rank of the diffusion tensor affects the ability of the Brownian particle 
to reach the equilibrium state.

\section{The stability of equilibrium state} \label{sec5}

In Ref.~\cite{cai2023relativisticII}, we assumed that the Brownian particle 
can always reach thermal equilibrium with the heat reservoir in the long time limit, 
without delving into the dynamical process for the Brownian particle to evolve 
from a non-equilibrium state to the equilibrium. To tackle this problem, 
the Lyapunov theorem offers a powerful tool.

For the sake of convenience,  we make an equivalent statement of the Lyapunov theorem: 
if there exists a functional for a given system that satisfies the condition of 
possessing at least one maximum point, with its time derivative strictly positive 
except at these maximum points, then the system is deemed asymptotically stable 
\cite{Teschl2012}. This functional is commonly referred to as the Lyapunov functional. 
The maximum points of the Lyapunov functional represent the stable states of the system. 
In the event that a system in a stable state experiences minor disturbances, 
as long as these disturbances do not surpass the local minimum of the Lyapunov 
functional, the system will return to its original stable state. Notably, 
if the maximum point of the Lyapunov functional is unique, no matter which 
initial state the system starts from, it will always reach the stable state. 
Such a property is called global asymptotically stable.

\begin{figure}[h]
\centering
\includegraphics[scale=.8]{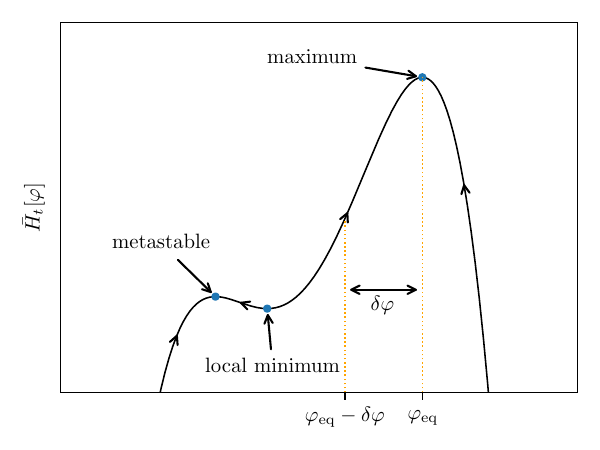}
\caption{The curve for an artificial Lyapunov functional. 
In the case with a full-rank diffusion 
tensor, our discussion rules out the possibility for the Lyapunov functional of
the charged relativistic Brownian 
particle to exhibit metastable and critical (i.e. inflection) points.}
\label{fig1}
\end{figure}

In our case, the expectation value of the relative entropy
\begin{align}
\bar H_t[\varphi]&:=\int_{\Sigma_{t}} \eta_{\Sigma_{t}} 
\dfrac{Z_{\mu}p^{\mu}}{m}\varphi\log\dfrac{\varphi}{\varphi_{\text{eq}}}
\end{align}
plays the role of Lyapunov functional of the Brownian particle. 
We now try to impose a small perturbation to the distribution function and calculate 
the variation of the above functional. The effect of the perturbation is to 
drive the system from one distribution to another, so the variation 
only acts on the distribution function. The result of the variation reads
\begin{align}\label{variation-H}
\delta \bar H_t[\varphi]&=\int \eta_{\Sigma_{t}} \dfrac{Z_{\mu}p^{\mu}}{m}
\delta\left[\varphi\(\log\varphi+\alpha-\beta_{\nu}p^{\nu}\)\right]\nonumber\\
&=\int \eta_{\Sigma_{t}} \dfrac{Z_{\mu}p^{\mu}}{m}
\left[\(1+\log\varphi+\alpha-\beta_{\nu}p^{\nu}\)\delta\varphi
+\varphi^{-1}(\delta\varphi)^{2}+O((\delta\varphi)^{2})\right]\nonumber\\
&=\int \eta_{\Sigma_{t}} \dfrac{Z_{\mu}p^{\mu}}{m}
\left[\(\log\varphi+\alpha-\beta_{\nu}p^{\nu}\)\delta\varphi
+\varphi^{-1}(\delta\varphi)^{2}+O((\delta\varphi)^{2})\right],
\end{align}
where, in the third line, the particle conservation has been used. 
It is obvious that the first order variation is zero and the second order variation 
is negative at the equilibrium state, so the equilibrium state is 
a maximum point of $\bar H_t[\varphi]$. Furthermore, the second order variation 
is not only negative at the equilibrium state but also at any states.
This implies that no critical points like the one illustrated in Fig.\ref{fig1} 
could appear in such systems, and hence the relative entropy of the Brownian particle 
is globally convex with only one maximum. 

Using eq.\eqref{time_derivative} and the second law, the time derivative of 
$\bar H_t[\varphi]$ can be confirmed to be non-negative 
\begin{align}
  \frac{\D}{\D t}\bar H_t[\varphi]=\int_{\mathcal S_t}\eta_{\mathcal{S}_t}\nabla_\mu H^\mu[\varphi]|\nabla t|^{-1}\ge 0.
\end{align}
However, the Lyapunov theorem requires it to be strictly positive except at the 
equilibrium state. According to the discussion made in the end of the last section, 
if the diffusion tensor is full-rank, the above inequality can only be saturated 
in the equilibrium state. Therefore, $\bar H_t[\varphi]$ is a Lyapunov functional 
of the Brownian particle. Fig.\ref{fig1} illustrates the general behavior of 
a Lyapunov functional with the inclusion of an equilibrium state and an artificial 
metastable state. In such cases, it is possible to drive the system away from the 
equilibrium state into the metastable state or vice versa by appropriate 
perturbations. Now, assuming a full-rank diffusion tensor, the above possibility 
is excluded because of the global asymptotic stability. 
Therefore, we can draw the conclusion that the relativistic 
charged Brownian particle system exhibits a tendency to evolve towards an increase 
in relative entropy, and the eventual state of evolution is characterized by 
the unique equilibrium distribution.

However, if the diffusion tensor is not full-rank, there will be some states 
other than the equilibrium state with zero entropy production rate. The Lyapunov functional 
may stop growing when the Brownian particle enters such states, and the distribution 
function no longer evolves toward the equilibrium one. Therefore, in the situation 
that the diffusion tensor is not full-rank, the Brownian particle may not be able to 
reach the equilibrium state.

\section{Concluding remarks}
\label{sec6}

With the aid of covariant formalism of relativistic stochastic mechanics, the first law 
and the second law of general relativistic stochastic thermodynamics have been established
for a system of charged Brownian particle subjecting to an external electromagnetic field. 
\blue{The the first and the second laws in stochastic thermodynamics differ from the 
usual first and second laws in traditional thermodynamics in that the former hold for an 
individual particle subjecting to the thermal influence of a heat reservoir 
and are more intimately connected with the underlying stochastic, in contrast 
to the latter ones which hold only for a whole macroscopic system.}
The fully covariant formulation \blue{which we adopted} has assisted in elucidating 
numerous conceptual issues. 
The final discussion about the stability of the equilibrium 
distribution helps us to clarify that the equilibrium state of Brownian particle 
is the state with maximum entropy. 

As a potential application scenario of the generic construction, we expect that a 
charged Brownian particle moving in Reissner-Nordstr\"om spacetime may be an excellent 
example for which the theory is applicable. In doing so, the relationship between 
non-equilibrium statistical physics in curved spacetime and black hole thermodynamics 
might become an interesting subject of future study. 

On the other hand, we have discussed only the first and second laws in this work, 
but a deeper understanding about the origin of the irreversibility in curved spacetime 
still calls for further investigation. At present, the best approach for the origin of irreversibility
is based on various fluctuation theorems \cite{kurchan1998fluctuation,searles1999fluctuation,crooks1999entropy,hatano2001steady,
chernyak2006path,seifert2012stochastic}. However, fluctuation theorems have not yet been 
realized in the framework of general relativity. We hope to come back to
this subject in some upcoming works.

\appendix

\section{Currents on spacetime and their divergences}
\label{app.A}

Let $\mathscr{O}(x,p)$ be a scalar field on the mass shell bundle, 
then the expectation value of $\mathscr{O}$ observed by Alice reads 
\begin{align}\label{average}
\bar{\mathscr{O}}_t[\varphi]&:=-\int \eta_{\Sigma_{t}}\mathscr{Z}_{A} \mathscr{J}^{A}[\varphi]\mathscr{O}=-\int \eta_{\mathcal{S}_{t}} Z_{\mu}\int\eta_{(\Gamma^{+}_{m})_{x}}\frac{p^{\mu}}{m}\varphi\mathscr{O}.
\end{align}
Since $\mathscr{Z}=Z^{\mu}e_{\mu}$ is the unit normal vector of $\Sigma_{t}$ 
and $\mathscr{J}[\varphi]$ is the probability current, 
$-\mathscr{Z}_{A}\mathscr{J}^{A}$ is clearly the probability density on $\Sigma_{t}$. 
The momentum space integral in eq.\eqref{average} is simply a current on 
spacetime manifold,
\begin{align}
\mathscr{O}^{\mu}[\varphi]:=\int\eta_{(\Gamma^{+}_{m})_{x}} 
\frac{p^{\mu}}{m}\varphi \mathscr{O}.
\end{align}
Therefore, the expectation value can also be written as a surface integral 
of $\mathscr{O}^{\mu}[\varphi]$ on the configuration space,
\begin{align}
  \bar{\mathscr{O}}_t[\varphi]=-\int_{\mathcal{S}_t}\eta_{\mathcal{S}_t}Z_\mu\mathscr{O}^\mu[\varphi],
\end{align}
which implies that $\mathscr{O}^\mu[\varphi]$ is the density current of $\mathscr{O}$. 

Let $V$ be an arbitrary region in the spacetime manifold $\mathcal{M}$, 
and $\Gamma=\{(x,p)\in\Gamma^+_m|x\in V\}$ is the corresponding region 
in the mass shell bundle. Let $y^\mu$ be the unit normal vector of 
the boundary $\partial V$, the corresponding unit normal vector of $\partial\Gamma$ 
is then $\mathscr{Y}=y^\mu e_\mu$. By the Gauss theorem, we have
\begin{align}
\int_V\eta_{\mathcal{M}}\nabla_\mu\mathscr{O}^\mu[\varphi]
&=\int_{\partial V}\eta_{\partial V} y_\mu\mathscr{O}^\mu[\varphi]
=\int_{\partial\Gamma}\eta_{\partial\Gamma}\frac{y_\mu p^\mu}{m}\varphi\mathscr{O}
=\int_{\partial\Gamma}\eta_{\partial\Gamma}
\mathscr{Y}_A\mathscr{J}^A[\varphi]\mathscr{O}\notag\\
&=\int_{\Gamma}\eta_{\Gamma^+_m}\mathscr{J}^A[\varphi]\hat\nabla^{(\hat h)}_A\mathscr{O}=\int_V\eta_{\mathcal M}\int_{(\Gamma^+_m)_x}\eta_{(\Gamma^+_m)_x}\mathscr{J}[\varphi](\mathscr{O}).
\end{align}
Since the region $V$ is arbitrary, the integrand on the right hand side 
and the left hand side should be equal to each other. Hence, 
the divergence of $\mathscr{O}^\mu[\varphi]$ is
\begin{align}\label{divergence}
\nabla_\mu\mathscr{O}^\mu[\varphi]
=\int_{(\Gamma^+_m)_x}\eta_{(\Gamma^+_m)_x}\mathscr{J}[\varphi](\mathscr{O}).
\end{align}

In relativistic thermodynamic relation, the time derivative of a physical 
quantity is usually replaced by the divergence of the associated current. 
This can be achieved as follows. Let $\mathcal S_{t_1}$ and $\mathcal S_{t_2}$ be
two Cauchy surfaces and $V$ be the region enclosed by a cylindrical surface 
with $\mathcal S_{t_1}$ and $\mathcal S_{t_2}$ playing as the upper and bottom. 
Then, integrating eq.\eqref{divergence} over $V$, we get, by Gauss theorem and 
co-area formula\cite{nicolaescu2011coarea,negro2022sample}, the following 
equation,
\begin{align}
\bar{\mathscr{O}}_{t_2}[\varphi]-\bar{\mathscr{O}}_{t_1}[\varphi]
=\int_V\eta_{\mathcal M}\nabla_\mu\mathscr{O}^\mu[\varphi]
=\int_{t_1}^{t_2}\D t\int_{\mathcal S_t}\eta_{\mathcal{S}_t}
\nabla_\mu\mathscr{O}^\mu[\varphi]|\nabla t|^{-1}.
\end{align}
Therefore, the time derivative of the expectation value is the integral 
of the divergence of the current times $|\nabla t|^{-1}$ on the 
configuration space,
\begin{align}\label{time_derivative}
\frac{\D}{\D t}\bar{\mathscr{O}}_{t}[\varphi]
=\int_{\mathcal S_t}\eta_{\mathcal{S}_t}\nabla_\mu\mathscr{O}^\mu[\varphi]|\nabla t|^{-1}.
\end{align}

\section*{Acknowledgement}

This work is supported by the National Natural Science Foundation of China under the grant
No. 12275138.

\section*{Data Availability Statement} 

This research has no associated data. 

\section*{Declaration of competing interest}

The authors declare no competing interest.

\providecommand{\href}[2]{#2}\begingroup
\footnotesize\itemsep=0pt
\providecommand{\eprint}[2][]{\href{http://arxiv.org/abs/#2}{arXiv:#2}}


\end{document}